
\magnification=\magstep1
\looseness=2
\overfullrule=0pt
\hsize=6.5 truein
\vsize=9.0 truein
\baselineskip=10pt plus2pt
\parskip=6pt plus4pt minus2pt

\font\csc=cmcsc10
\font\bigrm=cmr10 scaled \magstep2
\def\lcdm{$\Lambda$CDM \ }
\def\gtwid{\mathrel{\raise.3ex\hbox{$>$\kern-.75em\lower1ex\hbox{$\sim$}}}}
\def\ltwid{\mathrel{\raise.3ex\hbox{$<$\kern-.75em\lower1ex\hbox{$\sim$}}}}

\def\lsim{\ltwid}

\def\Mhsun{~h^{-1}M_{\odot}\ }

\def\Mpch{h^{-1}{\rm Mpc}}
\def\kpch{h^{-1}{\rm kpc}}

\def\noi{\noindent}
\def\etal{{\it et al.}\ }
\def\M10{{\times 10^{10} M_{\odot}\ }}


\def\hMpc{\ h^{-1}\ {\rm Mpc}}
  \def\ps{\noindent\goodbreak
  \parshape 2 0truecm 16.2truecm 2truecm 14.2truecm}
  \def\ref#1;#2;#3;#4;#5.{\ps #1\ #2, {#3},\ {#4}, #5}
  \def\\{\hfil\break}
  
  \def\apj{Astrophys. J.}
  
  \def\apjs{Astrophys. J. Suppl.}
  \def\mnras{M.N.R.A.S.}

\newcount\qno  \qno=0
\def\Eqnew#1{\global\advance\qno by 1%
             \xdef#1{\number\qno}}
\def\Eqn#1{\eqno(#1)}
\def\nxt#1{\Eqnew{#1}\Eqn{#1}}


\centerline{\bigrm SMALL-SCALE POWER SPECTRUM AND CORRELATIONS}
\vskip .1in
\centerline{\bigrm IN \char'3CDM }
\vskip .1in

\centerline{\csc Anatoly Klypin}
\centerline{Department of Astronomy, New Mexico State University}
\centerline{Las Cruces, NM 88001; aklypin@nmsu.edu}

\medskip
\centerline{{\csc Joel Primack}}
\centerline{Physics Department, University of California, Santa Cruz,
CA 95064; joel@lick.ucsc.edu}

\medskip
\centerline{\csc Jon Holtzman}
\centerline{Department of Astronomy, New Mexico State University}
\centerline{Las Cruces, NM 88001; holtz@nmsu.edu}
\centerline{\it Received 1995 \hskip 5em ; accepted}

\vfill\eject
\centerline{ABSTRACT}
Cosmological models with a positive cosmological constant
($\Lambda>0)$ and $\Omega_0 < 1$ have a number of attractive
features. A larger Hubble constant $H_0$, which can be compatible with
the recent HST estimate, and a large fraction of baryon density in
galaxy clusters make them current favorites. Early galaxy formation
also is considered as a welcome feature of these models. But early
galaxy formation implies that fluctuations on few megaparsec scales
spent more time in the nonlinear regime, as compared with standard
Cold Dark Matter (CDM) or Cold + Hot Dark Matter (CHDM) models. As has
been known for a long time, this results in excessive clustering on
small scales. We show that a typical $\Lambda$CDM model with $H_0=70$
km/s/Mpc, $\Omega_0=0.3$, and cosmological constant $\Lambda$ such
that $\Omega_\Lambda \equiv \Lambda/(3H_0^2) =1-\Omega_0$, normalized
to COBE on large scales and compatible with the number-density of
galaxy clusters, predicts a power spectrum of galaxy clustering in
real space which is too high: {\it at least} twice larger than CfA
estimates (Park \etal 1994) and 3 times larger than APM estimates
(Baugh \& Efstathiou 1994) for wavenumbers $k=(0.4-1)h/{\rm Mpc}$.
This conclusion holds if we assume either that galaxies trace the dark
matter ($\sigma_8\approx 1.1$ for this model) or just that a region
with higher density produces more galaxies than a region with lower
density. The only way to reconcile the model with the observed power
spectrum $P(k)$ is to assume that regions with high dark matter
density produce fewer galaxies than regions with low
density. Theoretically this is possible, but it seems very unlikely:
X-ray emission from groups and clusters indicates that places with a
large density of dark matter produce a large number of galaxies.
Since it follows that the low-$\Omega$ \lcdm models are in serious
trouble, we discuss which \lcdm models have the best hope of surviving
the confrontation with all available observational data.

\noindent {\it Subject headings:} cosmology: theory --- dark
matter --- large scale structure of the universe ---
galaxies: clustering

\vfil\eject

\bigskip
\centerline{1. INTRODUCTION}

Models with density $\rho_0=\Omega_0\rho_{\rm crit}$ less then the
critical density $\rho_0 < \rho_{\rm crit} = 3H_0^2/8\pi G$ and a
cosmological constant $\Lambda \neq 0$ have become popular in recent
years for a number of reasons.  With an age of the universe $t_0 \geq
13$ Gy, such models allow a large Hubble constant $H_0=(60-80)$
km/sec/Mpc, which agrees with observational indications for large
Hubble constant (e.g. Freedman \etal 1994; Riess, Press \& Kirshner
1994), although the value of $H_0$ remains uncertain.  These models
can naturally explain how galaxy clusters could have a large fraction
(20\%--30\%; White \etal 1993; White \& Fabian 1995) of mass in
baryons without severe contradictions with amount of baryons
compatible with primordial nucleosynthesis (Walker \etal 1991; Krauss
\& Kernan 1994; Copi, Schramm, \& Turner 1995). (Note however that the
latest results on clusters, taking into account the higher masses from
gravitational lensing, decreases the cluster baryon problem for
$\Omega=1$ models; see e.g.  Squires et al. 1995.)  Low-$\Omega$ \lcdm
models also cure many other problems of the standard CDM model (e.g.
Efstathiou, Sutherland, \& Maddox 1990; Bahcall \& Cen 1992; Kofman,
Gnedin, \& Bahcall 1993; Cen, Gnedin, \& Ostriker 1993; Gnedin 1995b;
Park \etal 1994; Croft \& Efstathiou 1994).

If the Universe has $\Omega_0 < 1$ with $\Lambda=0$, the model is not
compatible with standard inflation. This means that we do not have an
explanation for such fundamental observed properties of the Universe
as its homogeneity.  Measurements of cosmic microwave background
anisotropies by COBE and other CMB experiments put severe constraints
on any possible inhomogeneity of the Universe.  While rather contrived
inflationary schemes do appear to be able to generate universes that
look from inside like open models (Sasaki et al. 1993; Bucher,
Goldhaber, \& Turok 1995; Linde \& Mehlumzian 1995), it remains
uncertain what spectrum of fluctuations to use for structure formation
in such models, and also whether they produce a sufficiently
homogeneous universe.  Thus, open models would solve some problems,
but could open more difficult questions. This is the reason why we
consider in the present paper flat cosmological models with
cosmological constant $\Lambda$: $\Omega_{\rm total}=\Omega +
\Omega_{\Lambda} =1$. Such models can be compatible with inflation
scenario in its most general form.  They also give a larger $t_0$ than
models with the same $\Omega_0$ with $\Lambda=0$.

Dynamics of cosmological models with $\Lambda$ constant were discussed
in many papers (e.g. Lahav \etal 1991; Kofman, Gnedin, \& Bahcall
1993).  Already the first N-body simulations (Davis \etal 1985,
Gramann 1988) revealed a problem with the model: the correlation
function is too steep and too large at small scales.  For example,
Efstathiou \etal (1990) found that the model with $\Omega_0=0.20$
nicely matches the APM angular correlation function $w(\theta)$ on
large angular scales ($\theta > 1^{\circ}$), but disagrees by a factor
of 3 with APM results on smaller scales. The disagreement was not
considered serious because ``the models neglect physics that is likely
to be important on small scales where $\xi(r) >>1$''.

 While it is true that one cannot reliably take into account all
physics of galaxy formation, it does not mean that we can manipulate
the ``galaxies'' in models without any restrictions. For example, one
of the ways out of the problem would be to assume that places with
high dark matter density somehow are less efficient in producing
galaxies.  Coles (1993) shows that if the dark matter density has a
gaussian distribution and the number density of galaxies is any
function of local dark matter density (``local bias''), then the
correlation function of galaxies cannot be flatter then the
correlation function of the dark matter. Because the correlation
function of dark matter in the $\Lambda$CDM model was already too
steep at small scales, we cannot satisfy APM $w(\theta)$ constraints
{\it simultaneously} at large and small scales with any local
bias. (So, we can reduce small-scale $w(\theta)$, but this will ruin
it on large scales.) Thus, in order to reconcile the model with
observations we need to appeal to nonlocal effects as used by Babul \&
White (1991) on small scales or by Bower \etal (1993) on large
scales. Those nonlocal effects can result from photoionization of ISM
by UV photons produced by quasars, AGNs, and young galaxies. Another
source of non-locality is propagation of shock waves produced by
multiple supernovae in active galaxies (Ikeuchi \& Ostriker 1986). It
should be noted that both effects are not very efficient in
suppressing star formation in high density environments. UV radiation
heats gas only to few tens of thousands of degrees. This does affect
the formation of small galaxies and delays the time of formation of
the first stars on large galaxies, but it is difficult to see how can
it change a large galaxy. If the gas falls into the gravitational
potential of normal-size galaxy with effective temperature of about
$10^6$K, it does not matter much if it was ionized and preheated.
Even if a strong shock is produced by an active galaxy, it is
difficult to deliver the shock to a nearby galaxy. Because galaxies
are formed in very inhomogeneous environments, a shock produced by one
galaxy will have a tendency to damp its energy into local void, not to
propagate into a dense area where another galaxy is forming. Numerical
hydro+N-body simulations of Gnedin (1995a), which incorporate effects
of UV radiation, star formation, and supernovae explosions, do not
show any antibias of luminous matter relative to the dark matter.

In this paper we are trying to avoid complicated questions about
effects of star formation. We estimate the nonlinear power spectrum of
dark matter and reinforce the old conclusion that it is not compatible
with observed clustering of galaxies. Thus, the model which has a
formal bias parameter $b\sim 1$ must have extra (anti-) bias. Then we
put lower limits on possible correlation function of galaxies for any
local bias, which assumes that larger dark matter density results in
more galaxies. The lower limit on the power spectrum is 2--3 times
higher than CfA estimates (Park \etal 1994) and 3--4 times higher than
APM results (Baugh, \& Efstathiou 1994).

\bigskip
\centerline{2. MODEL AND NORMALIZATION}

We have chosen the following model as the first representative of
$\Lambda$CDM models to simulate: $\Omega_0=0.3$,
$\Omega_{\Lambda}=0.7$, $\Omega_{\rm bar}=0.026$, $h=0.70$. The age of
the Universe for the model is $t_{\rm Univ}=13.4$Gyr. The
normalization of the spectrum of fluctuations for the model
corresponds to the quadrupole of cosmic microwave anisotropy $Q_{\rm
rms-PS}=21.8\mu K$ (Stompor, Gorski, \& Banday 1995). This gives the
rms mass fluctuation $\sigma_8$ for top-hat filter with radius $8h^{-
1}$Mpc equal to $\sigma_8 =1.10$. With this amplitude the model
predicts a bulk velocity $V_{50}=355$km/s for a sphere with
$50h^{-1}$Mpc.

Another way of expressing and fixing the normalization is the number
of rich galaxy clusters. White, Efstathiou \& Frenk (1993) estimate a
cluster abundance of about $n_{\rm Cl}=4\times
10^{-6}(h^{-1}$Mpc$)^{-3}$ for masses exceeding $M_{Cl} \equiv
4.2\times 10^{14}h^{-1}M_{\odot}$. Biviano \etal (1993) give a
slightly higher estimate for the same mass limit: $n_{\rm Cl}=6\times
10^{-6}(h^{-1}$Mpc$)^{-3}$. The $\Lambda$CDM model considered in this
paper predicts $n_{\rm Cl}=3.9\times 10^{-6}(h^{-1}$Mpc$)^{-3}$ for $M
> M_{Cl}$, which is close to these results. (For discussion of cluster
masses in different models see Borgani \etal 1995.)  Here we use the
Press-Schechter approximation with gaussian filter and
$\delta_c=1.50$. A top-hat filter with $\delta_c=1.68$ gives very
similar results. White, Efstathiou, \& Frenk (1993) found that their
results of N-body simulations of clusters are quite well matched by
the Press-Schechter approximation with these parameters. These results
disagree with estimates of the mass function given by Bahcall \& Cen
(1993), who find $n_{\rm Cl} =1.8\times 10^{-6}(h^{-1}$Mpc$)^{-3}$,
which is 2--3 times lower than numbers given by White, Efstathiou \&
Frenk and by Biviano \etal The number of clusters is extremely
sensitive to the amplitude of fluctuations. For example, if instead of
$\sigma_8 =1.10$ we would take $\sigma_8 =0.80$ (Kofman, Gnedin, \&
Bahcall (1993), Gnedin (1995b)), the number of clusters with
$M>M_{Cl}$ for our model would drop to the level $n_{\rm Cl}=4.5\times
10^{-7}(h^{-1}$Mpc$)^{-3}$, which makes it unacceptable. This also
means that there is no room in the model for effects like
gravitational waves or a small tilt of the spectrum to reduce the
amplitude at small scales in order to ease the problem with the power
spectrum and correlation function: lower amplitude on 1--10~Mpc scales
makes the model inconsistent with the number of galaxy clusters.

\bigskip
\centerline{ 3. SIMULATIONS}

In order to estimate the power spectrum in the nonlinear regime we ran
6 simulations using standard Cloud-in-Cell Particle-Mesh code (Kates
\etal 1990, Hockney \& Eastwood 1981). Six simulations were run with
very different box sizes and resolutions (see Table 1). The
simulations were started at $z=45$ and run till $z=0$ with constant
step in expansion parameter $\Delta a =0.003$. By comparing models
with different resolution we can estimate effects of the resolution on
the power spectrum in the nonlinear regime.

In order to check our estimates of the number of clusters in the model
we ran a simulation with a much larger box and lower amplitude. In
this case the box size was $200h^{- 1}$Mpc, the resolution was $\Delta
x=390h^{-1}$kpc. The amplitude $\sigma_8=0.9$ was chosen above the
value $\sigma_8=0.75-0.8$ used by Kofman, Gnedin \& Bahcall (1993),
and Gnedin (1995b), but below our amplitude $\sigma_8=1.1$. We found
that the number density of ``clusters'' with mass larger than
$M=4.2\times 10^{14}h^{- 1}M_{\odot}$ within Abell radius
$r_A=1.5h^{-1}$Mpc is $n_{\rm Cl} =1.6\times
10^{-6}(h^{-1}$Mpc$)^{-3}$. Just as expected, it is well below Biviano
\etal (1993) and White
\etal (1993) estimates.

\bigskip
\centerline{4. RESULTS}

\bigskip
\centerline{\it 4.1 Power Spectrum}

Figure 1 presents the power spectrum of dark matter for our
\char'3CDM model. The bottom panel shows results of
different simulations. On small scales ($k > 1h$Mpc$^{-1}$) results
show definite convergence: better resolution leads to higher power
spectrum, but the difference between simulations gets smaller for
better and better resolution.  On scales $0.2h{\rm Mpc}^{-1}<k<1h{\rm
Mpc}^{-1}$ results are fluctuating because of cosmic variance (small
number of large structures in a simulation). The top panel shows the
averaged nonlinear power spectrum. The spectrum was averaged over all
simulations in the range $0.1h{\rm Mpc}^{- 1}<k<1h{\rm Mpc}^{-1}$. For
larger wavenumbers we used results from the two simulations with the
best resolution.  The spectrum matches the linear spectrum at $k\sim
0.2h{\rm Mpc}^{-1} $, but goes significantly above it at larger
wavenumbers.

In Figure 2 we compare the nonlinear power spectrum with observational
results. Dots are results of the Baugh \& Efstathiou (1993) for the
APM Galaxy survey. Open circles show results for the CfA survey (Park
\etal 1993). Formal error bars for each of the surveys are smaller
than the difference between the results.  For the CfA catalog the
power spectrum was estimated in red-shift space and then a correction
was made for velocity distortions (see Park \etal 1993 for details).
Because the corrections are large and model dependent (corrections
were found by comparing redshift-space and real-space power spectra of
dark matter in a large-box, low resolution \lcdm PM simulation), these
CfA results are slightly less reliable then the APM results. The CfA
catalog is also more shallow and it is possible that high values of
$P(k)$ in CfA are due to a few nearby large structures like the Great
Wall. The full curve represents the power spectrum of the dark matter
from Figure 1 (top panel). At $k=0.5h{\rm Mpc}^{-1}$ it is 3 times
larger than the APM estimate and 1.6 times larger than the CfA
value. Note that no simple scale independent bias can help to
reconsile this spectrum with observations because its shape is wrong.

We can place stronger constraints on the model by assuming that the
number of galaxies produced in some volume is related to the local
dark matter density: $n_{\rm gal}=f(\rho_{dm})$, where $f$ is some
function. One can interpret $n_{\rm gal}$ as a probability to produce
a galaxy given the density of dark matter $\rho_{dm}$. One also can
naively expect that $f$ is a monotonically growing function: the
larger the density, the bigger is the number of galaxies. This also
agrees with the fact that in places where we know there are large
amounts of dark matter (as indicated for example by X-ray emission
from clusters and groups), there are more galaxies. (In well-studied
clusters such as Abell 2218, the galaxies, gas indicated by X-rays,
and dark matter indicated by gravitational lensing or other methods
all have essentially the same profile; cf. Squires et al. 1995.)  As
an extreme case, we can assume that $f=constant$: if there is enough
mass to produce a galaxy, it does not matter what is the density. This
is {\it not} a reasonable assumption for galaxy formation. We do
observe that clusters and groups are the places with higher
concentration of galaxies as compared with, say, peripheral parts of
clusters or filaments. But this sets a limit on the possible power
spectrum of galaxies in this model: if anything, the power spectrum
will be larger than our estimate.

Numerically this was realized in the following way. Using our
simulations with the best resolution \char'3CDM$_c$ and
\char'3CDM$_f$ we construct the density field. Then density
in all cells  (cell sizes are given in Table 1) are set
to zero if the mass in the cell is less than $5\times
10^9M_{\odot}$. This is too small to produce galaxies for
APM or CfA catalogs. Even if we take a mass-to-light ratio
10, the absolute magnitude would be $M_V=-16.7$ --- only a
tiny fraction of galaxies with this magnitude is in the
catalogs. We assume that all cells with higher density could
produce galaxies. But following our arguments, all of them
are assigned constant density. The value of the constant
does not affect the result because we estimate the power
spectrum of fluctuations: $n_{\rm gal}(x)/\langle n_{\rm
gal} \rangle$.

Results for the power spectrum obtained in this way are shown in
Figure 2 as the dashed curve (\char'3CDM$_f$ simulation) and the
dot-dashed curve (\char'3CDM$_c$ simulation). Both models agree on
large scales, but on smaller scales ($k>0.5h{\rm Mpc}^{-1}$) the
higher resolution in \char'3CDM$_f$ resulted in a higher power
spectrum. Both models indicated that galaxies must be more clustered
than the dark matter at least on scales for which we have
observational data: $0.1h{\rm Mpc}^{-1}<k<1h{\rm Mpc}^{-1}$.
Comparison with APM and CfA results indicate contradictions on the
level of factor 2--4. A more conventional galaxy identification method
(e.g., high peaks) would imply larger discrepancies.

\bigskip
\centerline{\it 4.2 Correlation Function}

Correlation functions provide another way of looking at the same
problem. In Figure 3 we plot the real-space correlation function of
dark-matter particles $\xi_{\rm dm}$ as a full curve. For comparison
we also show the predictions of linear theory (triangles). The
dark-matter correlation function is an average of three simulations
with best resolution (\char'3CDM$_{c,f}$). A small correction was made
to take into account waves longer than the box size. If $\xi_{\rm
sim}$ is the correlation function in a simulation with the size of the
box $L$, then the corrected correlation function $\xi$ is $$ \xi(r) =
\xi_{\rm sim}(r) +\Delta\xi(r), \nxt{\Eqxi}
$$
\noindent where
$$
\Delta\xi(r) ={1\over 2\pi^2}\int_o^{2\pi/L}P(k){\sin(kr)\over kr}k^2dk.
         \nxt{\Eqxid}
$$
\noindent For radii $r$ less than $10h^{-1}$Mpc the correction only
slightly depends on $r$. For box size $L=80h^{-1}$Mpc it was
$\Delta\xi =0.145$ at $r=10h^{-1}$Mpc and $\Delta\xi =0.154$ at
$r<1h^{-1}$Mpc. For box size $L=50h^{-1}$Mpc the corrections were
0.344 and 0.393 correspondingly. The corrections are not important for
radii less than 3--4~Mpc.

 On scales
larger than $4h^{-1}$Mpc the  correlation function of the dark-matter
is only slightly higher than what is predicted by the linear theory.
But on smaller scales the difference is soaring: at $1h^{-1}$Mpc the
non-linear correlation function is 8 times higher than the linear
theory prediction. For $0.8\Mpch< r < 4\Mpch$ the correlation function
can be approximated by the power law:
$$
	\xi_{\rm dm}(r)=\left({6.5\Mpch\over r}\right)^{2.4}. \nxt{\Exidm}
$$
\noindent The slope 2.4 of the correlation function is too steep as
compared with the traditional 1.8.

Figure 3 also shows the usual power law approximation
$(5\Mpch/r)^{1.8}$ (dot-dashed line) and real-space results of the
Stromlo-APM survey (Loveday \etal 1995, Figure 2b) (asterisks).  The
errors of the latter are about 10\% for all points (the size of
markers on the plot) except for the first data point (smallest $r$),
where it is about 20--30\%.  It is interesting to note that on scales
1--5 $h^{-1}$ Mpc there is a very good match of the Stromlo-APM
correlation function and the correlation function given by the linear
theory (i.e., the fourier transform of the linear power spectrum
$P(k)$). In other words, in this cosmological model {\it there is no
room for nonlinear effects} , which are essential on $\ltwid 5\Mpch$
scale. In order for the model to survive, galaxies in the model must
be severely antibiased with the biasing parameter $b\approx 1/3$ at
$r=1\Mpch$.

In an effort to mimic galaxies in the model, we identified all dark
matter halos with $\delta\rho/\rho>200$ (see below) in our three
high-resolution simulations (\char'3CDM$_{c,f}$). The correlation
function of the halos is shown as the dashed curve in Figure 3. The
slope of the correlation function of halos is even steeper than that
of the correlation function of the dark matter for $0.5\Mpch< r <
2\Mpch$. Its amplitude at $1\Mpch$ has not changed. There is small
anticorrelation (relative to the dark-matter) on large scales
$r>3\Mpch$, which is consistent with linear bias $b_{\rm
lin}=1/\sigma_8=1/1.1$. The agreement with the Stromlo-APM data has
not improved on $\approx 1\Mpch$ scale, and is worse on smaller
scales.

The halos were identified in the following way:

\noi {\bf (1)} We started with the
density field defined on our $800^3$ mesh. All local density maxima
above overdensity limit 30 were tagged.  This provides a very large
list of candidates, which is used to construct a much shorter list of
more realistic halos.

\noi {\bf (2)}
We then found positions of density maxima independently of the mesh.
For that we placed a sphere of $70\kpch$ radius (our resolution)
around each tagged maximum. Centers of mass of all dark matter
particles within each sphere were found. Then sphere centers were
displaced to the centers of mass and the procedure was iterated until
convergence.

\noi {\bf (3)} After that only halos with maximum overdensity
larger than 200 were selected for analysis. The number of selected
halos was a factor of ten smaller than the number of tagged maxima.
The radius with mean overdensity 200 was found and this radius and
mass within this radius were assigned as the radius and mass of the
halo.

\noi {\bf (4)} Because the procedure of finding all neighbors within large
radius is very cpu-time consuming, we set a limit of $0.9\Mpch$ on the
maximum possible radius of a halo. Fewer than a dozen extremely large
halos were affected by the limit. But those few are very important
because they are clusters of galaxies. As the result of the limit,
peripheral parts of clusters may be underrepresented in our halo
``catalogs''. The situation is unclear. While because of overmerging
the central parts of clusters ($r\lsim 300\kpch$) is basically
structureless with only {\rm one} halo found by the algorithm, halos
are found in very large numbers in peripheral areas of clusters. The
typical number of halos with mass larger than $10^{11}M_{\odot}$
within $0.9\Mpch$ radius for a cluster is about 100. It might be too
small as compared with real clusters. This also indicates that we
might be missing some halos even outside the radius. For example, we
might be missing halos which already have fallen through the cluster
and were destroyed by its tidal field. Because the destruction of
halos depends on many details (rate of infall, age of clusters,
trajectories of halos, densities of halos and clusters), it is
difficult to estimate how important this effect is outside the
$0.9\Mpch$ radius.

\noi {\bf (5)} The destruction of halos in groups and clusters of galaxies
(``overmerging'') significantly affects the correlation function of
halos.  Note that destruction is not just due to the lack of
resolution (Moore, Katz, \& Lake 1995). It happens because dark matter
halos, which move in a cluster, have densities smaller than the
density of the halo at the center of the cluster. Any halo that comes
close to the center is distroyed by the tidal force.  We expect that,
in reality, galaxies survive the destruction because gas loses energy
and sinks to the center raising the density.  But this physics is of
course not included in dissipationless simulations.  In order to
overcome this problem we assume that big halos with mass $M$ above
some limit $M_{\rm bu}$ actually represent not a galaxy, but many
galaxies. Halos above $M_{\rm bu}$ are broken up into $M/M_{\rm bu}$
``galaxies'', which are then randomly distributed inside the big halo
in such a way that their number density falls down as $r^{-2}$ from
the center. We have chosen $M_{\rm bu}=7\times 10^{12}\Mhsun$, but
results are not sensitive to the particular value of $M_{\rm bu}$. In
order to avoid the problem of assigning different masses to broken
halos, we simply weight the contribution of every halo into the
correlation function by the mass of the halo. Thus the contribution of
a large halo to the correlation function at distances larger than
$1\Mpch$ is not affected at all by the breaking algorithm.

We have tried another, very different approach to the ``galaxy''
identification. It goes in line with what we did for the power
spectrum. We tagged dark matter particles with estimated overdensity
at the position of the particle above some limit. The density was
constructed using our $800^3$ mesh. The correlation functions of those
``galaxies'' is shown in Figure 4 as a long-dashed curve (overdensity
larger than 200), short-dashed curve (overdensity larger than 500),
and dotted curve (overdensity larger than 1000). Results represent
averages over the same three simulations considered above. For
comparison we also show $\xi_{\rm dm}$ (the full curve), Stromlo-APM
results (asterisks), and the 1.8 power law (dot-dashed line).  On
small scales the correlation function rises with rising overdensity
limit.  Surprisingly, we found no trend with the overdensity on scales
larger than $1\Mpch$. The curve for overdensity 200 is almost the same
(within 20 per cent) as the curve for halos (Figure 3).

Comparison of Figures 3 and 4 shows that results are very insensitive
to details of galaxy identification. It seems that none of the
simplest and most attractive schemes for the distribution of galaxies
in the model can give correlation functions that agree with
observations. Galaxies cannot follow the dark matter. Neither the
power spectrum (\S4.1) nor the correlation function allow this. The
simplest biasing models (halos with overdensity above 200 or density
above any reasonable threshold) do not work either: the discrepancy on
2--3~Mpc scales can be reduced, but the situation on smaller scale
gets even worse. We see three possible solutions for the contradiction
between observed and predicted correlation functions at $\sim 1\Mpch$
scale:

\noi {\bf (1)} On scales less then $\sim 1\Mpch$ galaxies  are ten times less
clustered than dark matter particles. This cannot be achieved by any
local bias. In order for the model to succeed, galaxy formation should
be significantly suppressed in galaxy clusters. It seems that there is
no observational indication of that: the mass-to-light ratio in
clusters does not differ much from that in groups.  Also the galaxy
distribution follows that of the dark matter.

\noi {\bf (2)} Estimates of the correlation function and the power spectrum
are significantly wrong. At $k=1h$Mpc$^{-1}$ the power spectrum in APM
and CfA surveys was underestimated by a factor 2--4, and the
Stromlo-APM correlation function is ten times too small at
$(1-2)\Mpch$.

\noi {\bf (3)} The model predicts clustering at $\sim 1\Mpch$ scale
which is not compatible with that in the real Universe.

The first alternative, while still possible, does not look very
attractive. If we assume it, we would need to admit that there is
almost no correlation of dark matter density and number density of
galaxies in high density areas. We would basically need to ``paint''
galaxies on small scales. This must be done carefully to avoid any
disturbances on large scales, which seem to be in accord with
observations, and also to avoid wrecking the hierarchical scaling
observed for galaxies (e.g. reduced skewness $S_3$ and curtosis $S_4$
are independent of scale $R$, where $S_n(R) =\langle
\xi^{(n)}\rangle_R/\langle_R \xi^{(2)}\rangle^{n-1}$; this is
non-trivial since the $n$-point correlations scale with the bias as
$\propto b^n$, thus leading to $S_n\propto b^n/b^{2n-2}=b^{2-n}$) (see
e.g. Bonometto et al. 1995 and refs. therein; cf. also Frieman \&
Gaztanaga 1994).  The second alternative also seems unlikely. While
errors are always possible, it seems that the APM and Stromlo-APM
surveys must have almost no relevance to the real distribution of
galaxies in order for the model to be viable. The third possibility
seems to be most likely. It does not mean that all variants of
\char'3CDM model are at fault, but it implies that the most
attractive variants with large age of the Universe, large Hubble
constant, and relatively large cosmological constant are very
difficult to reconcile simultaneously with the observed
clustering of galaxies and with the number of galaxy
clusters at the same time.

\bigskip
\centerline{5. DISCUSSION, AND ALTERNATIVE \lcdm MODELS}
\bigskip

It is instructive to compare the $\Omega_0=0.3$, $h=0.7$
\lcdm model that we have been considering with standard CDM
and with CHDM --- which is just CDM with the addition of some light
neutrinos whose mass totals about 5-7 eV (Klypin et al. 1995,
hereafter KNP95; Primack et al. 1995; and refs.  therein).  At $k=0.5
h$ Mpc$^{-1}$, Figs. 5 and 6 of KNP95 show that the $\Omega_\nu=0.3$
CHDM spectrum and that of a biased CDM model with the same
$\sigma_8=0.67$ are both in good agreement with the values indicated
for $P(k)$ by the APM and CfA data, while the CDM spectrum with
$\sigma_8=1$ is higher by about a factor of two.  In the present paper
we see that for the \lcdm model we have simulated, $P(k)$ for the same
$k$ is at least as high as CDM, and even higher for larger $k$.
Unless the data is misleading, or some sort of complicated galaxy
formation physics leads to a $P(k)$ below our estimates, this is a
serious problem for the \lcdm and standard CDM models.

But the \lcdm model we have considered here is but one of many.  So
let us look in $h$-$\Omega_0$ space for alternatives.  In Table 2 we
have compared predictions of a number of such models with data on
several length scales.  In this Table, all the models (except for the
\lcdm one marked with an asterisk) are normalized to the 2-year COBE
data as recommended by Stompor, Gorski, and Banday (1995).  The first
two lines of numbers give our estimates of several observational
quantities and the uncertainties in them, from large to small scales.
The bulk velocity at $r=50\hMpc$ is derived from the latest POTENT
analysis (Dekel 1995 and private communication); the error includes
the error from the analysis but not cosmic variance.  However, similar
constraints come from other data on large scales such as power spectra
that may be less affected by cosmic variance since they probe a larger
volume of the universe.  We have estimated the current number density
of clusters ($N_{\rm clust}$) with $M > 10^{15} h^{-1} M_\odot$ from
comparison of data on the cluster temperature function from X-ray
observations with hydrodynamic simulations (Bryan et al. 1994) as well
as from number counts of clusters (White et al. 1993, Biviano et al.
1993). All recent estimates of the cluster correlation function give
fairly large values at $30 \hMpc$ (e.g. Olivier et al. 1993, Klypin \&
Rhee 1994, and refs.  therein); this also suggests that the zero
crossing of the correlation function must exceed $\sim 40 \hMpc$.

The cluster constraint is powerful.  Models with Hubble constant $h$
as high as 0.8 cannot have $\Omega_0$ larger than 0.2 and still
satisfy the age constraint $t_0>13$ Gy, but for such high $h$ and low
$\Omega_0$ the COBE normalization leads to too low a cluster density.
The cluster density is in good agreement with observations for the
$\Omega_0=0.3$, $h=0.7$ \lcdm model that we have considered in this
paper, but as we mentioned above this means that we do not have the
freedom to lower the normalization in order to lower the value of the
nonlinear $P(k)$.  Table 2 shows that the linear $P(k)$ is already
fairly high at $k=0.1h$ Mpc$^{-1}$, and as Fig. 2 shows, the nonlinear
spectrum is considerably higher than the linear one, especially at
larger $k$ where the conflict with the APM and CfA data becomes
significant.  Of course, as we lower $h$ and raise $\Omega_0$, we
approach standard CDM for which the cluster abundance is far too high
with COBE normalization.  Thus for \lcdm models with intermediate
$\Omega_0$ and $h$ correspondingly low to satisfy the $t_0$
constraint, the cluster abundance goes up.  Consider for example
$\Omega_0=0.5$ and $h=0.6$.  With COBE normalization the cluster
abundance is about five times too high, which means that there is room
to lower the spectrum and perhaps resolve the $P(k)$ and $\xi(r)$
discrepancies on small scales without exceeding the COBE upper limit
on large scales.  For example, the line in the table marked
$\Lambda$CDM* represents a biased version of the same model, with the
spectrum simply lowered by a constant factor.  (This could in
principle be due to gravity waves contributing on the COBE scale,
although we admit that we are unaware of an inflationary model with a
large gravity wave contribution but no tilt.  Indeed, recent work on
inflationary cosmologies in the context of supersymmetric models
(Dine, Randall, and Thomas 1995; Ross and Sarkar 1995) suggests that
inflation will occur at a lower energy scale so that gravity wave
production will not be cosmologically significant.  Alternatively, we
could tilt the model, i.e.  assume a primordial spectrum $P_p(k)
\approx Ak^n_p$ with primordial spectral index $n_p$ less than the
Zel'dovich value of unity.  Or we could consider a $\Lambda$CHDM
model, i.e. \lcdm with a small contribution to the density from light
neutrinos.  The case shown in Table 2 corresponds to a single neutrino
species with a mass of 2.44 eV, consistent with the preliminary
results from the LSND experiment (Athanassopoulos et al. 1995).  All
of these alternatives lower the spectrum significantly and would no
doubt result in $P(k)$ in better agreement with the data than the
COBE-normalized $\Omega_0=0.3$, $h=0.7$ \lcdm model considered here.
The same is true of the CHDM models, represented in Table 2 by two
variants of the two-neutrino model with $\Omega_\nu=0.20$.  The first
of these is the Harrison-Zel'dovich ($n=1$) version considered in
Primack et al. 1995, but with the new COBE normalization; now the
cluster density is too high.  But as the next line shows, a small tilt
($n=0.96$) with gravity waves corresponding to a $V \propto \phi^2$
chaotic inflationary model gives an excellent fit to all the data
considered in the Table. All these models would differ in their
predictions for other measurable quantities such as small-angle CMB
anisotropies in the first Doppler peak region, production of early
objects such as damped Lyman $\alpha$ systems, and small-scale
velocities.  We regard the further investigation of all these models
as an important project for the future.

Acknowledgments.  We thank Stefano Borgani for helpful comments on an
earlier draft of this paper.  This work was partially supported by NSF
grants at NMSU and UCSC.  Simulations were performed on the CM-5 and
Convex 3880 at the National Center for Supercomputing Applications,
University of Illinois, Champaign-Urbana, IL.

\vfill\eject
\centerline{ REFERENCES}
\medskip
\parindent =0pt
\parskip =0pt

\ps    Athanassopoulos et al. 1995, Phys. Rev. Lett. submitted
\ps    Babul, A., \& White, S.D.M. 1991, \mnras, 250, 407
\ps    Bahcall, N.A., \& Cen, R. 1992, \apj, 398, L81
\ps    Bahcall, N.A., Cen, R., 1993, \apj, 407, L49
\ps    Baugh, C.M., \& Efstathiou, G. 1994, \mnras, 267, 323
\ps    Biviano, A., Girardi, M., Giuricin, G., Mardirossian, F., \&
	Mezzetti, M. 1993, \apj, 411, L13
\ps    Borgani, S., Plionis, M., Coles, P., \& Moscardini, L. 1995,
	\mnras, in press
\ps    Bonometto, S., Borgani, S., Ghigna, S., Klypin, A., \& Pri\-mack, J.
	 1995, \mnras, 273, 101
\ps    Bower, R.G., Coles, P., Frenk, C.S., \& White, S.D.M. 1993, \apj,
	405, 403
\ps    Bryan, G., Klypin, A., Loken, C., Norman, M., \& Burns, J 1994,
	\apj, 437, L5
\ps    Bucher, M., Goldhaber, A.S., \& Turok, N. 1995,
	hep-ph/9411206
\ps    Cen, R., Gnedin, N.Y. \& Ostriker, J.P. 1993, \apj, 417, 387
\ps    Coles, P. 1993, \mnras, 262, 591
\ps    Copi, C.J., Schramm, D., \& Turner, M.S. 1995 Science,
       267, 192
\ps    Croft, R.A., \& Efstathiou, G. 1994, \mnras, 267, 390
\ps    Dekel, A. {\it Ann. Rev. Astron. Astroph.} 1994, 54, 1,
	 and private communications
\ps    Dine, M., Randall, L., \& Thomas, S. 1995,  hep-ph/9503303
\ps    Davis, M., Efstathiou, G., Frenk, C.S., \& White, S.D.M. 1985,
	\apj, 292, 371
\ps    Efstathiou, G., Sutherland, W.J., \& Maddox, S.J. 1990, Nature,
       348, 705
\ps    Freedman, W. \etal 1994, Nature, 371, 757
\ps    Frieman, J., \& Gaztanaga, 1994, \apj 
\ps    Gnedin, N. 1995a, \apjs, 97, 231
\ps    Gnedin, N. 1995b, \apj, in press
\ps    Gramann, M. 1988, \mnras, 234, 569
\ps     Hockney, R.W., \& Eastwood, J.W. 1981, {\it Numerical
        simulations using particles} (New York: McGraw-Hill)
\ps    Ikeuchi, S., \& Ostriker, J.P. 1986, \apj, 301, 522
\ps    Kates, R.E., Kotok, E.V., \& Klypin, A.A. 1991, A\&A, 243, 295
\ps    Klypin, A.,  \& Rhee, G. 1994, \apj,  428, 399
\ps    Klypin, A.,  Nolthenius, R., \& Primack. J. 1995, \apj, submitted
\ps    Kofman, L.A., Gnedin, N.Y., \& Bahcall, N.A. 1993, \apj, 413, 1
\ps    Krauss, L.M., \& Kernan, P.J. 1994, \apj, 432, L79
\ps    Lahav, O., Rees, M.J., Lilje, P.B., \& Primack, J. 1991, \mnras,
	251, L128
\ps    Linde, A., \&  Mehlumzian, A. 1995, astro-ph/9506017
\ps    Loveday, J., Efstathiou, G., Maddox, S.J., \& Peterson, B.A. 1995,
	\apj, in press
\ps    Moore, B., Katz, N., \& Lake, G. 1995, preprint.
\ps    Olivier,S., Primack, J., Blumenthal, G.R., \&  Dekel, A. 1993,
	\apj,  408 17
\ps    Park, C.,Vogeley, M.S., Geller, M.J. \& Huchra, J.P.
       1994, \apj, 431, 569
\ps    Primack, J., Holtzman, J., Klypin, A., \& Caldwell, D. 1995, Phys. Rev.
	Lett., 74. 2160
\ps    Riess, A.G., Press, W.H., \& Kirshner, R.P. 1995, \apj, 438, L17
\ps    Ross, G. G., \& Sarkar, S. 1995, hep-ph/9506283
\ps    Sasaki, M., Tanaka, T., Yamamoto, K., \& Yokoyama, J. 1993,
	Phys. Lett. B317, 510
\ps    Squires, G., Kaiser, N., Babul, A., Fahlman, G., Woods, D.,
	 Neumann, M., \&  B\"ohringer H. 1995,  astro-ph/9507008
\ps    Stompor, R., Gorski, K.M., \& Banday, L. 1995, \mnras, in press
\ps    Walker, T.P., Steigman, G., Schramm, D.N., Olive, K.A., \& Kang,
	H.S. 1991, \apj, 376, 51
\ps    White, S.D.M., Efstathiou, G., \& Frenk, C.S. 1993,
	\mnras, 262, 1023
\ps    White, S.D.M., Navarro, J.F., Evrard, A.E., \& Frenk, C.S. 1993,
	Nature, 366, 429
\ps    White, D.A., \& Fabian, A.C. 1995, \mnras, 273, 72
\vfill
\eject

\centerline{FIGURE CAPTIONS}
\bigskip

\noi Figure 1. Power spectrum of dark matter for \char'3CDM
model. The bottom panel shows results of different simulations. On
small scales ($k > 1h$Mpc$^{-1}$) results show definite convergence:
better resolution leads to higher power spectrum, but the difference
between simulations gets smaller for better and better resolution. On
scales $0.2h{\rm Mpc}^{-1}<k<1h{\rm Mpc}^{-1}$ results are fluctuating
because of cosmic variance (small number of large structures in a
simulation). The top panel shows the averaged power spectrum.

\noi Figure 2.  Comparison of the nonlinear power spectrum in
\char'3CDM model with observational results. Dots are results for the
APM Galaxy survey. Open circles show results for the CfA
survey. Formal error bars for each of the surveys are smaller than the
difference between the results. The full curve represents the power
spectrum of the dark matter from Figure 1 (top panel).

\noi Figure 3.  Real-space correlation
function of the dark-matter (full curve) in \char'3CDM model. For
comparison we also show predictions of the linear theory (triangles).
The usual power law approximation $(5\Mpch/r)^{1.8}$ is shown as the
dot-dashed line.  Real-space results of the Stromlo-APM survey are
presented by asterisks. Correlation function of halos with overdensity
more than 200 is shown by the dashed curve.

\noi Figure 4. The correlation function of dark matter with
overdensity more than 200 (log-dashed curve), 500 (short-dashed
curve), and 1000 (dotted curve). Other curves and symbols in this
Figure are the same as in Figure 3. Note that while the correlation
function on small scales ($<1\Mpch$) significantly depends on the
density threshold, it does not depend on the threshold on large
scales.

\bye